# Acoustic Measurement of Potato Cannon Velocity

## Michael Courtney and Amy Courtney

This article describes measurement of potato cannon velocity with a digitized microphone signal. A microphone is attached to the potato cannon muzzle and a potato is fired at an aluminum target about 10 m away. The potato's flight time can be determined from the acoustic waveform by subtracting the time in the barrel and time for sound to return from the target. The potato velocity is simply the flight distance divided by the flight time.

A potato cannon[1] is a simple device constructed from PVC pipe and a lantern sparker that uses aerosol hairspray as a propellant to launch potatoes 200 m or more. The principle of operation is similar to that of pistons in an engine and projectiles in a firearm. A spark ignites the fuel-air mixture that burns rapidly to produce hot expanding gases that propel the potato out of the barrel.

There are a number of treatments of potato cannons and similar pressure operated devices.[2] A lot of physics can be demonstrated with these devices, but it would be advantageous to have an independent determination of projectile velocity to confirm the mechanics presented. Velocity estimates can be inferred from range, but these usually depend on neglecting air resistance. This approach is no more accurate for a potato whose range is 100-200m than it is for a home-run baseball. We present an acoustic method for determination of potato velocity that should be accurate to better than 5%.[3]

A microphone is attached to the end of the barrel. An aluminum target is placed at a carefully measured distance roughly 10 m away. Targets with resonating sounds like a gong work better than targets with non-resonant, "clink" type of sounds. It works well to use a 6 mm thick aluminum sheet with its base corners resting on two blocks of wood and one point on a vertical edge leaning against a saw horse.

The sound waveform can be captured using a Vernier LabPro with microphone or a PC soundcard with software (such as Audacity[4]) that allows viewing the sound waveform to determine the time difference between the ignition of the propellant and the target sound. The total time from the sound of the propellant ignition to the sound of the target strike is the sum of the barrel time ($t_{barrel}$), the flight time ($t_{flight}$), and the time for the sound to return from the target to microphone ($t_{sound}$):

$$t_{total} = t_{barrel} + t_{flight +} \ t_{sound}.$$

The time for sound to travel back from the target to the microphone is the distance ($d$) divided by the velocity of sound ($V_{sound}$),

$$t_{sound} = d \ / \ V_{sound}.$$

$V_{sound}$ is roughly 331 m/s, but it varies with temperature approximately according to

$V_{sound}$= *331* m/s + *0.6* (m/s°C) *T*,

where *T* is the temperature in Celsius. In addition to having an accurate distance measurement, one needs the ambient temperature within 1°C.

The time in the barrel is the barrel length (*l*) divided by average velocity in the barrel. If the potato accelerated at a constant rate in the barrel, this would be exactly half of the potato flight velocity ($V_{potato}$). Acceleration in the barrel is not constant, but increases as the propellant burns and then decreases slightly after the propellant has been consumed. $V_{potato}$ */3* is a reasonable estimate of the average velocity in the barrel so that the total time of flight can be written as

$$t_{total} = \frac{3l}{V_{potato}} + \frac{d}{V_{potato}} + t_{sound}.$$

Solving for $V_{potato}$ yields

$$V_{potato} = \frac{3l + d}{t_{total} - t_{sound}}.$$

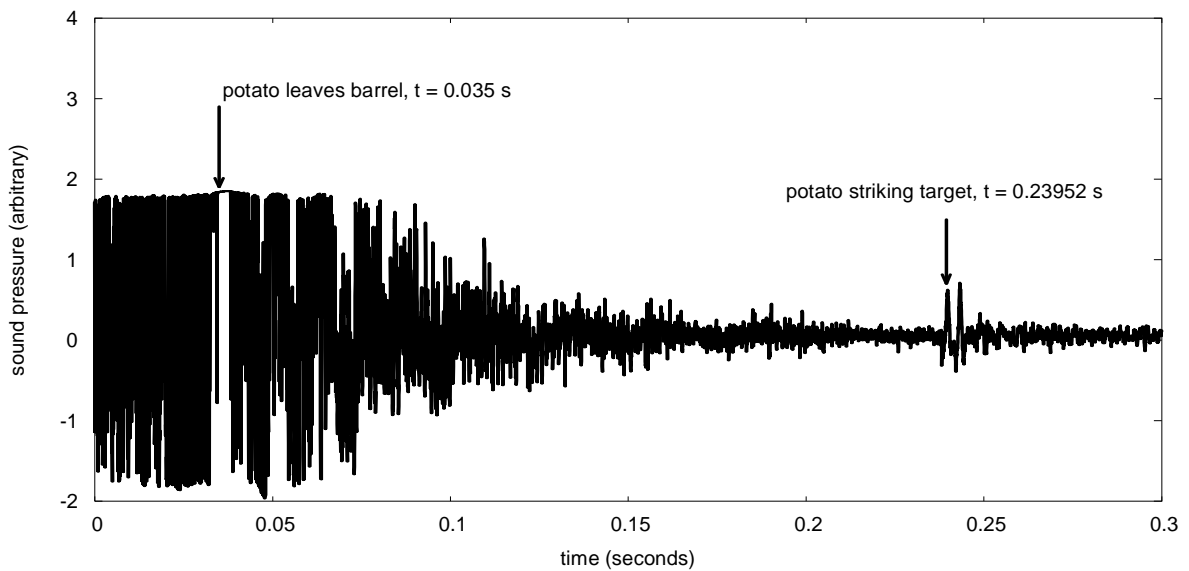

Figure 2 shows a sound waveform for the potato cannon fired at the aluminum target. The data acquisition was set to trigger on the propellant ignition, so this occurs at *t = 0 s*. The time of the target strike is $t_{total}$ *= 0.23952s.*

The speed of sound at the ambient temperature of 6°C is 334.6 m/s. Consequently, the time for the sound to return from target to microphone over the distance *d = 10.13 m* is

$t_{sound} = 10.13$ m/(334.6 m/s) = 0.03027 s,

so that the time between ignition and target strike is 0.20925 s.

Using the barrel length of 0.775m and substituting into the equation for potato velocity gives $V_{potato}$ = 59.52 m/s. Note that this potato flight velocity implies an average barrel velocity of $V_{barrel}$=19.84 m/s suggesting a barrel time of $t_{barrel}$ = 0.0391 s which is in fair agreement with the time assigned by identifying the gap of time when the pressure is pegged high in Figure 2. (When the potato leaves the barrel, the microphone is hit with the high pressure gasses escaping from the barrel.)

This validates our approximation that the average velocity of the potato in the barrel is one third of the flight velocity, $V_{potato}$ /3. Alternatively, if we use the time when the microphone signal is saturated by the escaping gases as the time the potato leaves the barrel, we obtain a flight velocity of $V_{potato}$ = 58.13 m/s, which is 2.3% lower than the value obtained estimating the barrel velocity as 1/3 of the flight velocity.

In summary, a potato cannon and a digitized microphone system allow for the measurement of potato flight velocity.

**About the Authors:**

**Michael Courtney** earned a PhD in Atomic, Molecular, and Optical Physics from MIT, worked for 7 years as an RF test engineer for Cisco Systems, and has taught college physics and forensic science for five years. He currently serves as the Director of the Forensic Science Program at Western Carolina University. *Western Carolina University, Cullowhee, NC 28723;* Michael_Courtney@alum.mit.edu

**Amy Courtney** earned a PhD in Medical Engineering/Medical Physics from Harvard University and has worked as a research scientist for Reebok, Intl. and the Cleveland Clinic. She is a founding member of the Ballistics Testing Group, which is developing new methods in terminal ballistics and in the acoustic reconstruction of shooting events. Ballistics Testing Group, P.O. Box 3103, Cullowhee, NC 28723. Amy_Courtney@post.harvard.edu